\documentclass[aps, twocolumn,showpacs,superscriptaddress,pra]{revtex4}
\usepackage{amssymb}
\usepackage{amsmath}
\usepackage{graphicx}
\usepackage{epsfig}
\usepackage{dcolumn}% Align table columns on decimal point
\usepackage{bm}% bold math
\usepackage{bbm}
\usepackage{color}
\usepackage{hyperref}
\usepackage[up]{subfigure}
\newcommand{\be}{\begin{equation}}
\newcommand{\ee}{\end{equation}}
\newcommand{\bc}{\begin{center}}
\newcommand{\ec}{\end{center}}
\newcommand{\bea}{\begin{eqnarray}}
\newcommand{\eea}{\end{eqnarray}}
\newcommand{\ra}{\rangle}

\begin{document}
\title{Quantum walk on distinguishable non-interacting many-particles and indistinguishable two-particle}
\author{C. M. \surname{Chandrashekar}}
\email{cmadaiah@phys.ucc.ie}
\affiliation{Physics Department, University College Cork, Cork, Ireland }
\author{Th.\surname{Busch}}
\email{thbusch@caesar.ucc.ie}
\affiliation{Physics Department, University College Cork, Cork, Ireland  }
%==================================================
\begin{abstract}
We present an investigation of many-particle quantum walks in systems of non-interacting distinguishable particles.  Along with  a redistribution of the many-particle density profile we show that the collective evolution of the many-particle system resembles the single-particle quantum walk evolution when the number of steps is greater than the number of particles in the system.  For non-uniform initial states we show that the quantum walks can be effectively used to separate the basis states of the particle in position space and grouping like state together. We also discuss a two-particle  quantum walk on a two-dimensional lattice and demonstrate an evolution leading to the localization of both particles at the center of the lattice. Finally we discuss the outcome of a quantum walk of two indistinguishable particles interacting at some point during the evolution. 
\end{abstract}


%==================================================
\maketitle
%==================================================
%================Introduction===================
\section{Introduction}
%===============================================

The idea of a quantum walk, the quantum analog of the classical random walk, dates back to 1958 \cite{Ria58} and 1965 \cite{FH65} but the concept was formally developed only in
1990's \cite{ADZ93, DM96, FG98}. In a one-dimensional situation a quantum walk evolving in position space spreads quadratically faster than its classical counterpart, due to the interference of amplitudes of the multiple paths \cite{ABN01, NV01}. This was found to have interesting applications in quantum information theory, allowing for efficient quantum algorithms \cite{Amb03, CCD03, SKB03, AKR05}. However, quantum walks have also been shown the be useful for coherent  quantum control over atoms and quantum phase transitions \cite{CL08}, to explain breakdown phenomena in electric-field driven systems \cite{OKA05}, to give direct experimental  evidence  for  wavelike energy  transfer  within photosynthetic systems \cite{ECR07,  MRL08}, to generate entanglement between two spatially separated system \cite{CGB10} and to generate topological phases \cite{KRB10}.
Experimental  implementation of quantum walks have been reported using nuclear magnetic resonance (NMR) \cite{DLX03, RLB05}, continuous tunneling of light fields  through waveguide lattices \cite{PLP08}, the  phase space  of trapped  ions  \cite{SMS09, ZKG10}, single optically trapped neutral atoms  \cite{KFC09} and single  \cite{SCP10, BFL10} and two-photon systems \cite{PLM10, OBB11}. All of these advances have made the area of quantum walks a very promising tool, just like its classical counterpart.

Quantum walks are widely categorized into two forms, namely, continuous-time and discrete-time walks. In this article we will discuss the discrete-time quantum walk and in particular 
we will focus on the quantum walk in a system of distinguishable particles.  In Section \ref{mpqw1} we will define the distinguishable non-interacting many-particle quantum walk and discuss the dynamics and some of the results of the collective evolution of the many-particle system. We  will show that for an evolution in which the number of steps is greater than the number of particles, the collective probability distribution resembles the single-particle probability distribution. This can be efficiently used for separating the different  basis states of the many particle system and grouping them together in position space even when the initial states of the particles are a randomized superposition of state. We also discuss the physical relevance of the study. In Section \ref{2pqw}, we look at the dynamics of a two-particle quantum walk and show that it is possible to localize the joint probability at the center of the lattice. We also look into the joint probability  of the indistinguishable, both boson and fermion two-particle quantum walk evolution when the particle meet in the lattice after the walk evolution. We conclude in Section \ref{conc}.

Though the review articles in this special issue introduce the concepts of quantum walks in great detail, we will briefly discuss the main features of the quantum walk evolution which will be relevant for this article here for self-consistency reasons. 

The discrete-time quantum walk of a single two-state particle in one-dimension is defined on a Hilbert space ${\cal H} =  {\cal H}_c\otimes    {\cal     H}_p$, where ${\cal H}_c$ is the coin Hilbert space with the basis state described in terms of the internal state of the particle, $|\downarrow \rangle = \begin{bmatrix} 1  \\ 0 \end{bmatrix}$ and $|\uparrow \rangle =\begin{bmatrix} 0  \\ 1 \end{bmatrix}$. The position Hilbert space, ${\cal H}_p$, has the basis states $|\psi_j\rangle$, where $j \in {\mathbbm  I}$ is a set of integers associated with each lattice site. Each step in the evolution of the walk is described using a  quantum coin operation 
\be
\label{qw:coin}
B (\theta)      \equiv  \left(      \begin{array}{clcr}
  \cos(\theta)      &     &    \mbox{~~} \sin(\theta)
  \\ \sin(\theta) & &  -\cos(\theta) 
\end{array} \right)
\ee
which evolves the particle into a superposition of the internal basis states and which is followed by the unitary shift operator 
\be
S\equiv         \sum_j \left [  |\downarrow \rangle\langle
\downarrow |\otimes|\psi_{j-1}\rangle\langle   \psi_j|   +  | \uparrow \rangle\langle
\uparrow|\otimes |\psi_{j+1}\rangle\langle \psi_j| \right ]
\ee
which transforms the state of the particle into a superposition in position space. Therefore, the full operation for each step of the quantum walk on the Hilbert space ${\cal H}_c\otimes    {\cal     H}_p$ can be written in the form 
\be
\label{Wop}
W (\theta)\equiv S[B(\theta) \otimes  {\mathbbm 1}].
\ee
and the state after $t$ steps of evolution is given by
\be
|\Psi_t\rangle=W(\theta)^t|\Psi_{\rm in}\rangle.
\ee  
Here  
\be
\label{qw:in}
|\Psi_{\rm in}\rangle= \left ( \cos(\delta/2)| \downarrow \rangle + e^{i\eta}
\sin(\delta/2)|\uparrow \rangle \right )\otimes |\psi_{0}\rangle,
\ee
is the initial state of the particle at a position $j=0$. 
The coin parameter $\theta$ controls the variance of the probability distribution of the walk \cite{ABN01, NV01, Kon02, CSL08}
and the probability to find the particle at site $j$ after $t$ steps is given by $P(j,t)  = \langle
\psi_j|{\rm tr}_c (|\Psi_t\rangle\langle\Psi_t|)|\psi_j\rangle$.

%=================
\section{Distinguishable many-particle quantum walk}
\label{mpqw1}
%=================
Many particle quantum walks are fundamentally different for systems of
non-interacting distinguishable and indistinguishable particles. For
the first one, the evolution of the walk can be straight forwardly
predicted by considering many single-particle quantum walks
\cite{CL08, GC10}, whereas for the latter one many particle
interference effects, based on the bosonic or fermionic nature of the
particles, strongly influence the evolution and makes it
computationally hard to study \cite{MTM11, RSS11}.

\begin{figure}[ht]
\subfigure[]{\includegraphics[width=9.0cm]{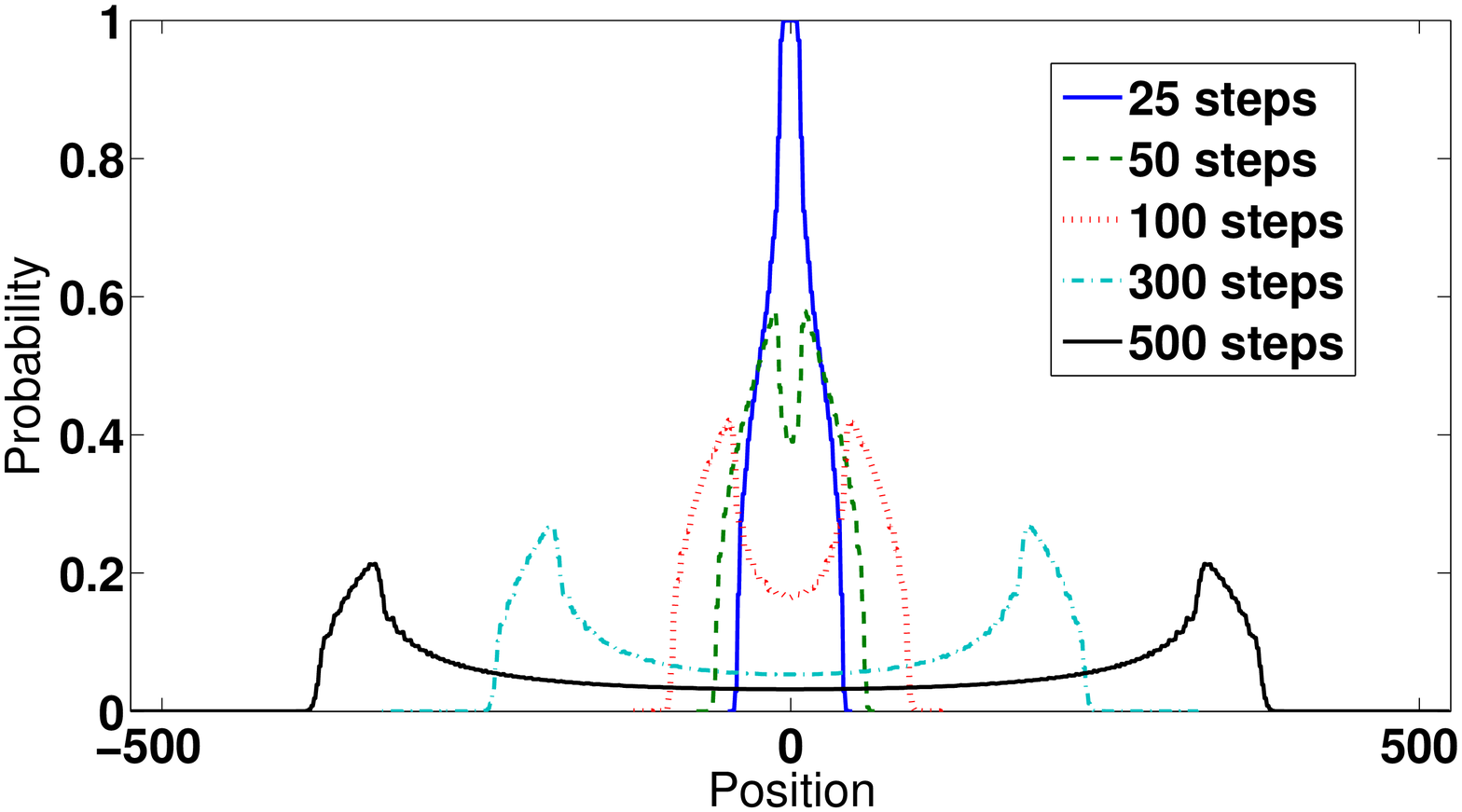}
\label{1a}}
\subfigure[]{\includegraphics[width=9.0cm]{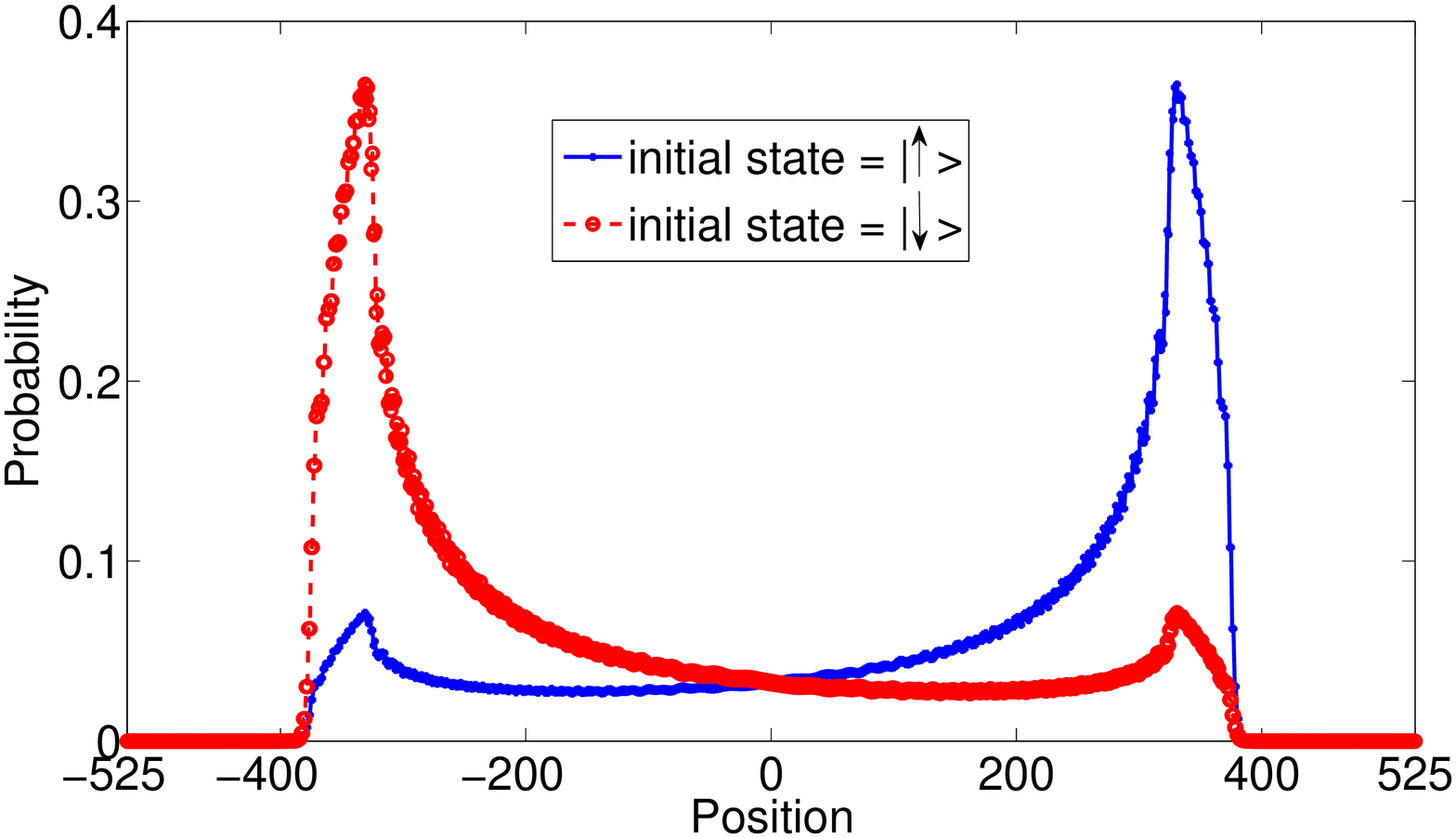}
\label{1b}}
\caption{(color online) Probability distribution of 51 particles initially located individually at positions $j = -25$ to $j = + 25$ after the quantum walk evolution using 
using the Hadamard operator $B( \pi/4)$  as the quantum coin.  (a) The initial state of all the  particles is $\frac{1}{\sqrt 2}(|\downarrow \ra + i | \uparrow \ra)$ and evolves in position space. The spread of the distribution with an increasing number of steps is clearly visible. (b) Initially all particles are either in state $|\downarrow \ra$  or state $| \uparrow \ra$  and subsequently subjected to the quantum walk of 500 steps.  For all the particle in the initial state $|\downarrow \ra$, the distribution with peak on the left is obtained and for all the particle in the initial state $|\uparrow \ra$, the distribution with peak on the right is obtained.} 
\end{figure}
\begin{figure}[ht]
\subfigure[]{\includegraphics[width=8.0cm]{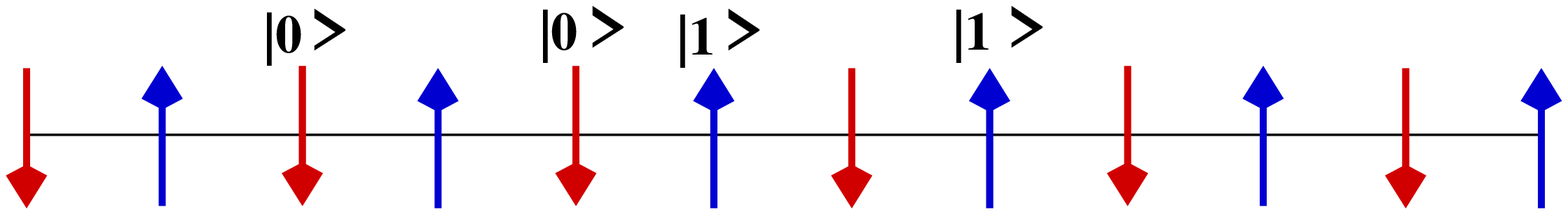}
\label{2a}}\\
\subfigure[]{\includegraphics[width=9.0cm]{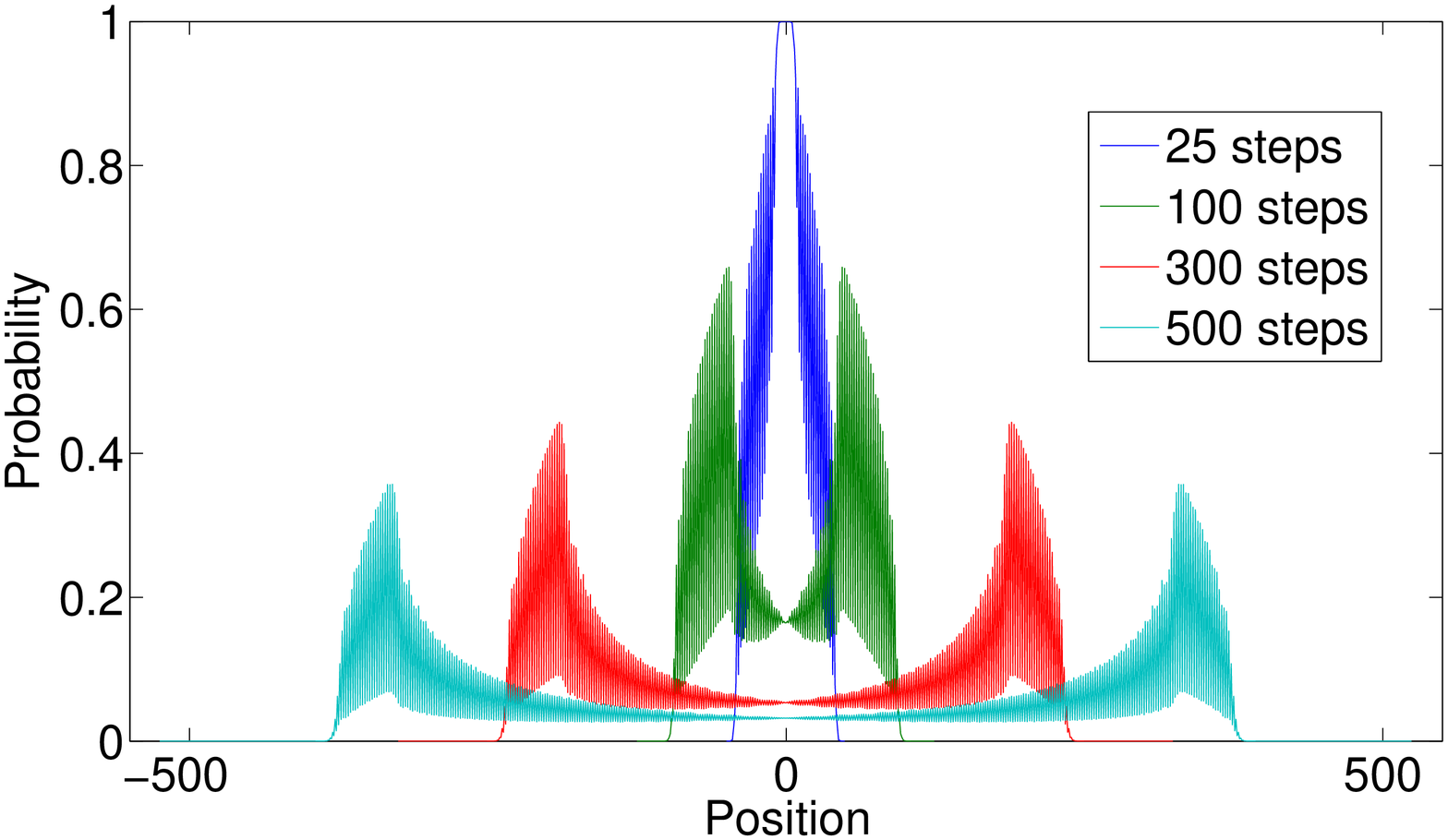}
\label{2b}}
\caption{(color online)(a) The initial state has an antiferromagnetic ordering with neighboring particles being in different internal states. (b) Probability distribution of 51 particles, initially in the state shown in (a) with one particle in each position ranging from $j = -25$ to $j = + 25$, after a quantum walk of different number of steps using the Hadamard operator $B( \pi/4)$  as the quantum coin.  Lower parts in the left (right) of the distribution are due to the low contribution of state $|\downarrow \rangle$ ( $|\uparrow \rangle$) from the particle initially in state $|\uparrow \rangle$ ($|\downarrow \rangle$).} 
\end{figure}
 \begin{figure}[ht]
\includegraphics[width=8.5cm]{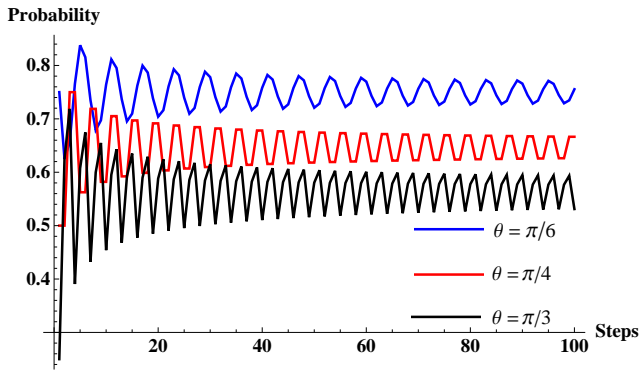}
\caption{(color online) Probability of finding a particle in state $|\downarrow \rangle$ of a single-particle walk with increasing number of steps if the particle was initially in state $|\downarrow\rangle$. For small values of $\theta$ the probability of finding the particle in state $|\downarrow \rangle$ is larger than for larger value of $\theta$. }
\label{3a}
\end{figure}
Though the evolution of distinguishable particles does not involve
many-particle interference effect, the collective behavior of the
single particle interference effects can reveal interesting features
of the systems dynamics. Such systems can be approximately realized in
cold, but thermal samples of neutral atomic gases in optical lattices
\cite{KFC09}, which can be engineered to minimize the atom-atom interaction and dynamically control the atom transport \cite{MGW03, DDL03, Jak04}.  
Therefore, they have been suggested for observation of quantum
phase transitions \cite{CL08} or for generation and control over
spatial entanglement between different lattice sites
\cite{GC10}. Another interesting question is the exploration of the
meeting probabilities and meting times of many-particles at pre-defined
positions (see Ref.\,\cite{SKJ06} for two particle meeting
probabilities).
 \par
In this section we will define the non-interacting distinguishable
many-particle quantum walk and discuss some of the interesting
outcomes from the collective evolution. To define a simple form of
distinguishable many-particle quantum walk in one-dimension, we will
consider a system of $M$ non-interacting particles, where initially
exactly one particle occupies a lattice site and every particle has
its own coin and position Hilbert space,
$\mathcal{H}=\left(\mathcal{H}_c \otimes \mathcal{H}_p
\right)^{\otimes M}$. If the number of particles is odd~\footnote{To
  have symmetry in labeling the position space we have chosen an odd
  number, however all results hold for even number of particles as
  well}, the initial state can be written as
\begin{equation}
  |\Psi_{ins}^{M}\rangle = \bigotimes_{j=-\frac{M-1}{2}}^{j=\frac{M-1}{2}} \left [
  \left( \frac{| \downarrow \rangle + i|\uparrow \rangle}{\sqrt{2}} \right) \otimes
  |\psi_{j}\rangle \right ],
  \label{initialMBQWstate}
\end{equation}
which, after $t$ steps, will evolve into  
\begin{equation}
  |\Psi_{t}^{M}\rangle =  [W(\theta)^{\otimes M}]^t \bigotimes_{j=-\frac{M-1}{2}}^{j=\frac{M-1}{2}}
  \left [ \left( \frac{| \downarrow \rangle + i| \uparrow \rangle}{\sqrt{2}} \right) \otimes
  |\psi_{j}\rangle \right ].
  \label{initialMBQWstate1}
\end{equation}
Here $W(\theta)^{\otimes M}$ is the evolution operator for each step of the walk, which will evolve each particle into the superposition of its neighboring positions, establishing the quantum correlation between the particle and the position space. After $t$ steps, these correlations overlap resulting in,
%\begin{widetext}
\begin{eqnarray}
	  \label{eq:manyUMI}
	  [W(\theta)^{\otimes M}]^{t}|\Psi_{in}^M\rangle \propto 
	   \bigotimes_{j = -\frac{M-1}{2}}^{\frac{M-1}{2}}  [  \mathcal{A}_{j-t}| \downarrow\rangle \otimes |\psi_{j-t}\rangle \nonumber \\
	  + \mathcal{A}_{j-t+1} |\downarrow \rangle \otimes |\psi_{j-t+1}\rangle +......+\mathcal{A}_{j+t}|\downarrow \rangle \otimes |\psi_{j+t}\rangle  \nonumber \\
+ \mathcal{B}_{j-t}|\uparrow \rangle \otimes |\psi_{j-t}\rangle + ......+ \mathcal{B}_{j+t})|\uparrow \rangle \otimes  |\psi_{j+t}\rangle  ],
\end{eqnarray}
which can be written as
\begin{eqnarray}
\label{eq:manyUMIa}
[W(\theta)^{\otimes M}]^{t}|\Psi_{in}^M\rangle \propto  \nonumber \\
 \bigotimes_{j = -\frac{M-1}{2}}^{\frac{M-1}{2}}\left( \sum\limits_{x=j-t}^{j+t} [ \mathcal{A}_x^j |\downarrow \rangle + \mathcal{B}_x^j |\uparrow \rangle ] \otimes |\psi_x\rangle \right ).
\end{eqnarray}
%\end{widetext}
Here $\mathcal{A}_x^j$ and $\mathcal{B}_x^j$ are the probability amplitudes of the state $|\downarrow \rangle$ and $|\uparrow \rangle$ of each of the particles initially at position $j$ at the new position $x$, which range from $(j-t)$ to $(j+t)$ after the $t$ step walk. 

The probability distribution after $t$ steps is given by the sum of the probabilities at a given lattice site of each particle $k$ 
\be
P(j, t) = \sum_{k=1}^{M}P_{k}(j, t)
\ee
and the effective probability distribution for different numbers of steps is shown in Fig. \ref{1a}. As expected, after $t$ steps of the quantum walk, the $M$ particles are spread between $(M-t)$ and $(M+t)$. In Fig. \ref{1b} the asymmetric probability distributions resulting from all particles being initially in state $|\downarrow \rangle$ or $|\uparrow \rangle$ after 500 steps are shown.  From Fig. \ref{1a}, and \ref{1b}, it is clearly evident that when $t\gg M$ the probability distribution profile of $M$ particles resembles the single-particle profile. From earlier studies of single-particle quantum walks of $t$ steps on a particle initially at position $j=0$ using $B_{\theta}$ as the quantum coin it is known that the probability distribution spreads over the interval $(-t \cos(\theta), t \cos(\theta))$ in position space and decays quickly outside this region \cite{NV01, CSL08}. For an $M$ particle system the peak of the effective probability distribution is given by contributions from the probability of all $M$ particles and therefore located at $\mp[t \cos(\theta) - M/2]$.
\par
Apart from all particles being initially in the symmetric superposition state of $|\downarrow \rangle$ and $| \uparrow \rangle$ (see Eq.(\ref{initialMBQWstate})), one can also consider a situation of antiferromagnetic ordering (see Fig. \ref{2a}), where two neighboring particles are in opposite states.  In Fig. \ref{2b} we show the final probability distribution for this situation after a different number of steps. Though each particle undergoes an asymmetric evolution with the states $|\downarrow \rangle$ moving left and the states $|\uparrow \rangle$  moving right, the collective distribution is symmetric due to equal number of particles initially in both states. 
\par
Many-particle quantum walks of particles initially in antiferromagnetic order or in a completely randomized initial state are very useful for separating different basis states of the particles in position space and grouping them together. Using a different angle $\theta$ in the quantum coin operation one can find different outcomes for the probabilities of basis states grouped after the evolution of the quantum walk.  To demonstrate this we will consider the examples of a single-particle initially in one of the basis state. 
%\begin{widetext}
\begin{figure}
\subfigure[]{\includegraphics[width=6.7cm]{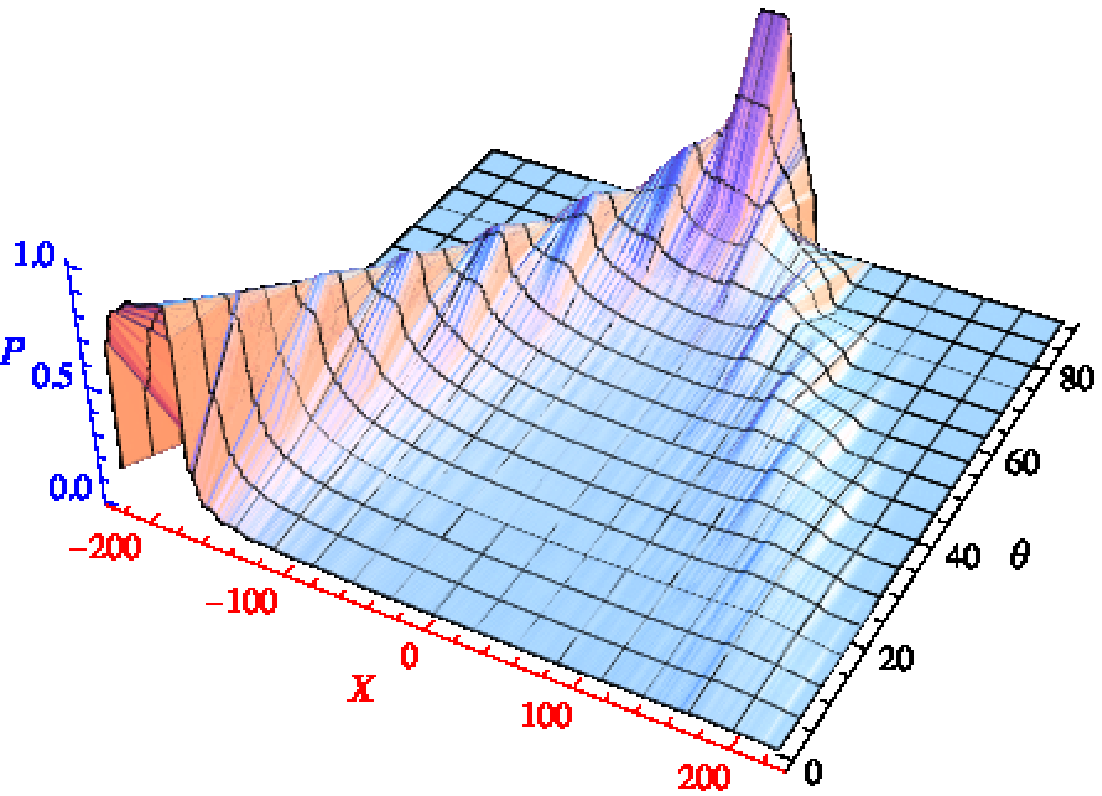}
\label{31a}}
\subfigure[]{\includegraphics[width=6.7cm]{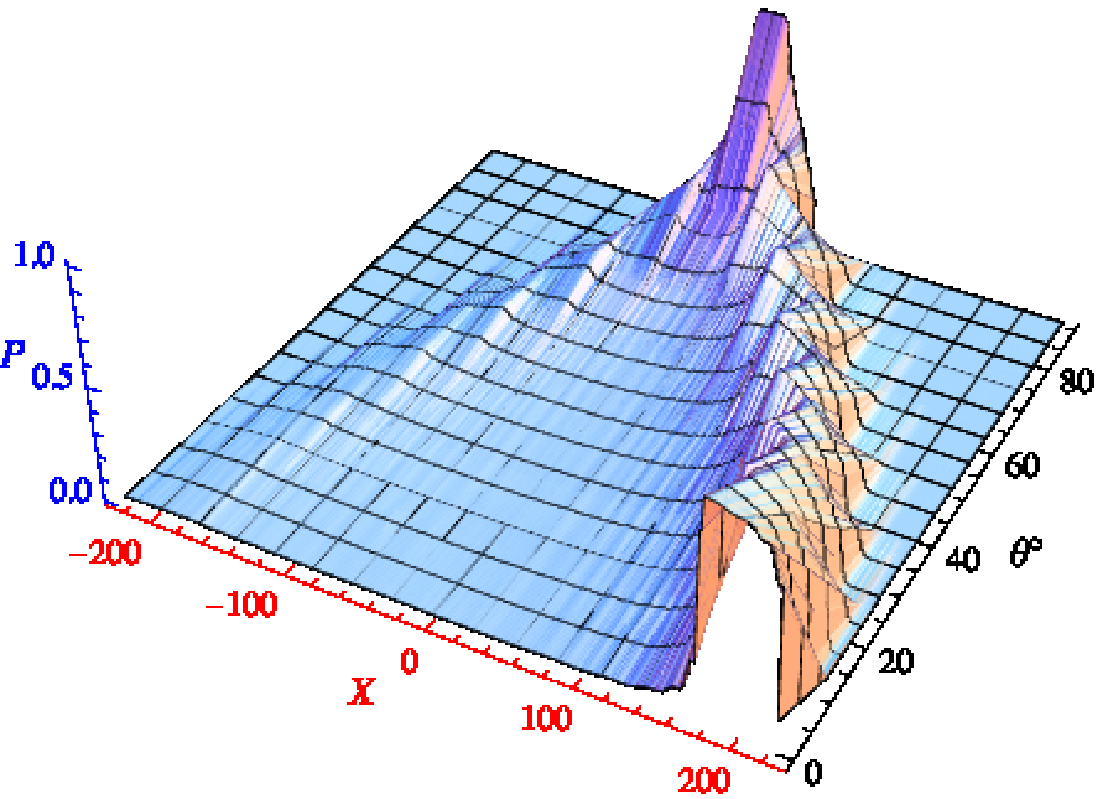}
\label{31b}}
\subfigure[]{\includegraphics[width=6.7cm]{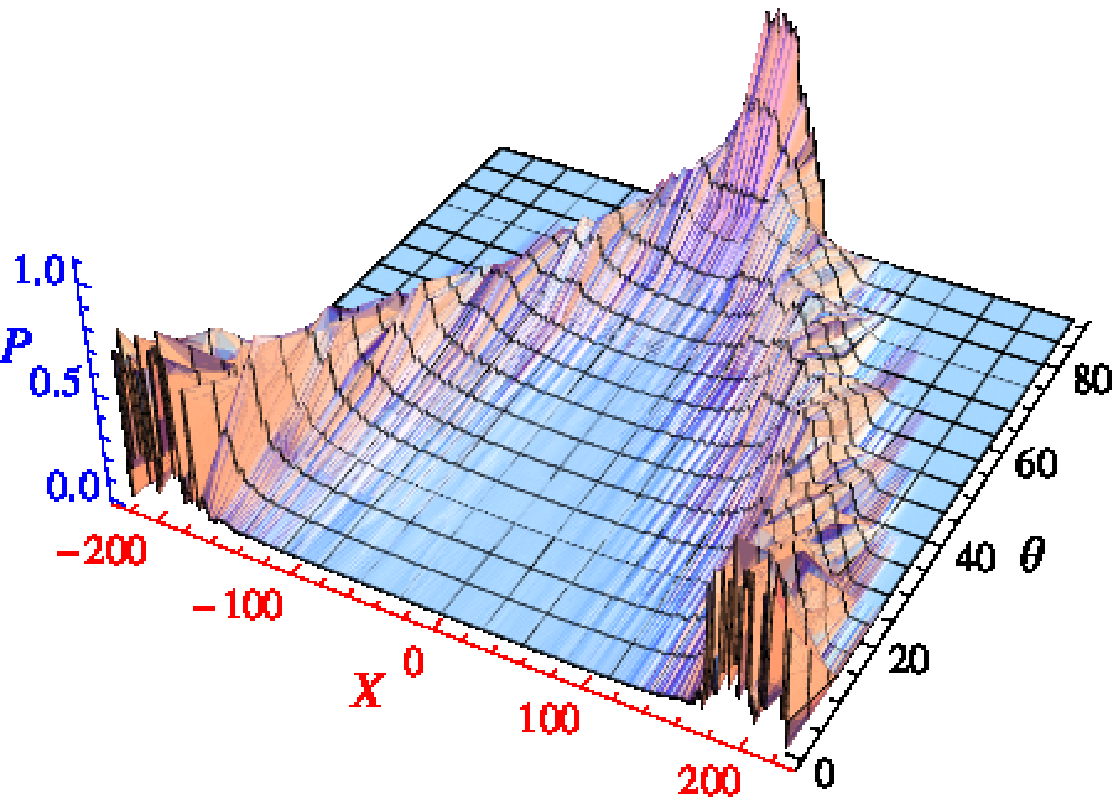}
\label{31c}}
\caption{(color online) Probability distribution of 51 particles after
  200 steps for different values of $\theta$. The states $|\downarrow
  \rangle$ walk to the left and the states $|\uparrow \rangle$ to the
  right. In (a) the initial state of all particle was $|\downarrow
  \rangle$ and in (b) the initial state of all particle was $|\uparrow
  \rangle$. In (c) the initial state of each particle is randomly
  choose from $|\downarrow \rangle$ and $|\uparrow \rangle$.}
\label{intqwe}
\end{figure}
%\end{widetext}
\par
A quantum walk of a single particle initially in state
  $|\downarrow \rangle$ (or equivalently $|\uparrow \rangle$) using a
  Hadamard coin ($\theta = \pi/4$) results in constructive
  interference towards the left (right) of the origin and therefore
  localizes all particles with high probability on the left (right) of
  the origin. To understand this, let us look at the analytic form
  of the evolution after $t$ steps using $B(\theta)$ as coin
  operator. The state after $t$ steps can be written as 
\begin{equation}
  W(\theta)^t |\Psi_{ins}\rangle = |\Psi(t)\rangle = \sum_{j=-t}^t 
  (\mathcal{A}_{j,t}|\downarrow \rangle|\psi_j\rangle +\mathcal{B}_{j,t}|\uparrow\rangle|\psi_j\rangle)
\label{eq:lr}
\end{equation}
where $\mathcal{A}_{j,t}$ and $\mathcal{B}_{j,t}$ are given by the
coupled iterative relations
\begin{subequations}
\label{eq:iter}
\begin{eqnarray}
\mathcal{A}_{j,t} = \cos(\theta) \mathcal{A}_{j+1,t-1} + \sin(\theta) \mathcal{B}_{j+1,t-1} \\
\mathcal{B}_{j,t} =-\cos(\theta) \mathcal{B}_{j-1,t-1} + \sin(\theta) \mathcal{A}_{j-1,t-1}.
\end{eqnarray}
\end{subequations}
Straightforward algebra allows to decouple these equations at the
price of a time-dependence on the previous two steps
\begin{subequations}
\label{eq:iter0}
\begin{eqnarray}
\mathcal{A}_{j,t}  =  \cos(\theta)\left (\mathcal{A}_{j+1,t-1} - \mathcal{A}_{j-1,t-1} \right ) - \mathcal{A}_{j, t-2}\\
\mathcal{B}_{j,t} = 
\cos(\theta)\left (\mathcal{B}_{j+1,t-1} - \mathcal{B}_{j-1,t-1} \right ) - \mathcal{B}_{j,t-2}.
\end{eqnarray}
\end{subequations}
By repeating this process of substitution one can find an expression
linking $\mathcal{A}_{j,t}$ and $\mathcal{B}_{j,t}$ to the amplitude
of the initial state of the particle and the angle, $\theta$, of the
coin operation.  Therefore, the expression for the total probability of
finding the particle in state $|\downarrow \rangle$ and
$|\uparrow\rangle$ after time $t$ is
\begin{subequations}
\label{eq:iter01}
\begin{eqnarray}
P_{|\downarrow\rangle} (t)= \sum_{j}|\mathcal{A}_{j,t}|^2 \\
P_{|\uparrow\rangle} (t)= \sum_{j}|\mathcal{B}_{j,t}|^2
\end{eqnarray}
\end{subequations}
To obtain a spatially symmetric probability distribution for a
particle initially in symmetric superposition state, the walk should
be invariant under an exchange of $|0\rangle \leftrightarrow
|1\rangle$, and hence should evolve $\mathcal{A}_{j,t}$ and
$\mathcal{B}_{j,t}$ alike (as, for example, the Hadamard walk does
\cite{KRS03}). From the above analysis we see that $\mathcal{A}_{j,t}$
and $\mathcal{B}_{j,t}$ are symmetric to each other and evolve alike
for all value of $\theta$ only when the initial state of the particle
is a symmetric superposition state. When the initial state is
$|\downarrow \rangle$, the walk will evolves with constructive
interference towards left and destructive interference to the right,
(exact form depending on the value of $\theta$) and vice versa when
the initial state is $|\uparrow \rangle$. The associated probability
amplitudes oscillate strongly between the left and the right hand side
for small numbers of steps and stabilize for longer times. This can be
seen in Fig.\,\ref{3a} and also directly from Eq.\,(\ref{eq:iter0})
when realizing that the amplitude at each position oscillates and the
range of oscillation reduces as the amplitude at each position
decreases over time \cite{Rom10}. 
\par
 For a particle initially in state $|\downarrow \rangle$ a smaller
value of $\theta$ returns a high probability of finding the particle
in state $|\downarrow \rangle$ but if the initial state is $|\uparrow
\rangle$ the probability of finding the particle in $|\downarrow
\rangle$ will be very low.  We should also note that a small
probability of state $|\uparrow \rangle$ ($|\downarrow \rangle$) is
present along with the state $|\downarrow \rangle$ ($|\uparrow
\rangle$) to the left (right) of the origin but that will not alter
the trend. From the above analysis we can conclude that an initially
randomized many-particle state can be efficiently sorted in position
space with respect to its basis states. This in turn allows to create
an ordered state with high probability.

To demonstrate this we show in Fig.\,\ref{31a} the probability
  distribution for different values of $\theta$ for a sample of 51
  particles after 200 steps when the initial state of all the
  particles was $|\downarrow \rangle$ and Fig.\,\ref{31b} shows the
  same for an initial state of $|\uparrow \rangle$. A strong asymmetry
  is visible for both cases. In contrast, the probability distribution
  shown in Fig.\,\ref{31c} assumes that the initial state of each
  particle was randomly chosen from $|\downarrow \rangle$ and
  $|\uparrow \rangle$ and the anisotropy in the final distribution
  vanishes. From Figs.\,\ref{31a} and (\ref{31b}) one can also see
  that for increasing $\theta$ the probability distribution widens and
  its maximum amplitude decreases.
  \begin{figure}[ht]
\subfigure[]{\includegraphics[width=6.6cm]{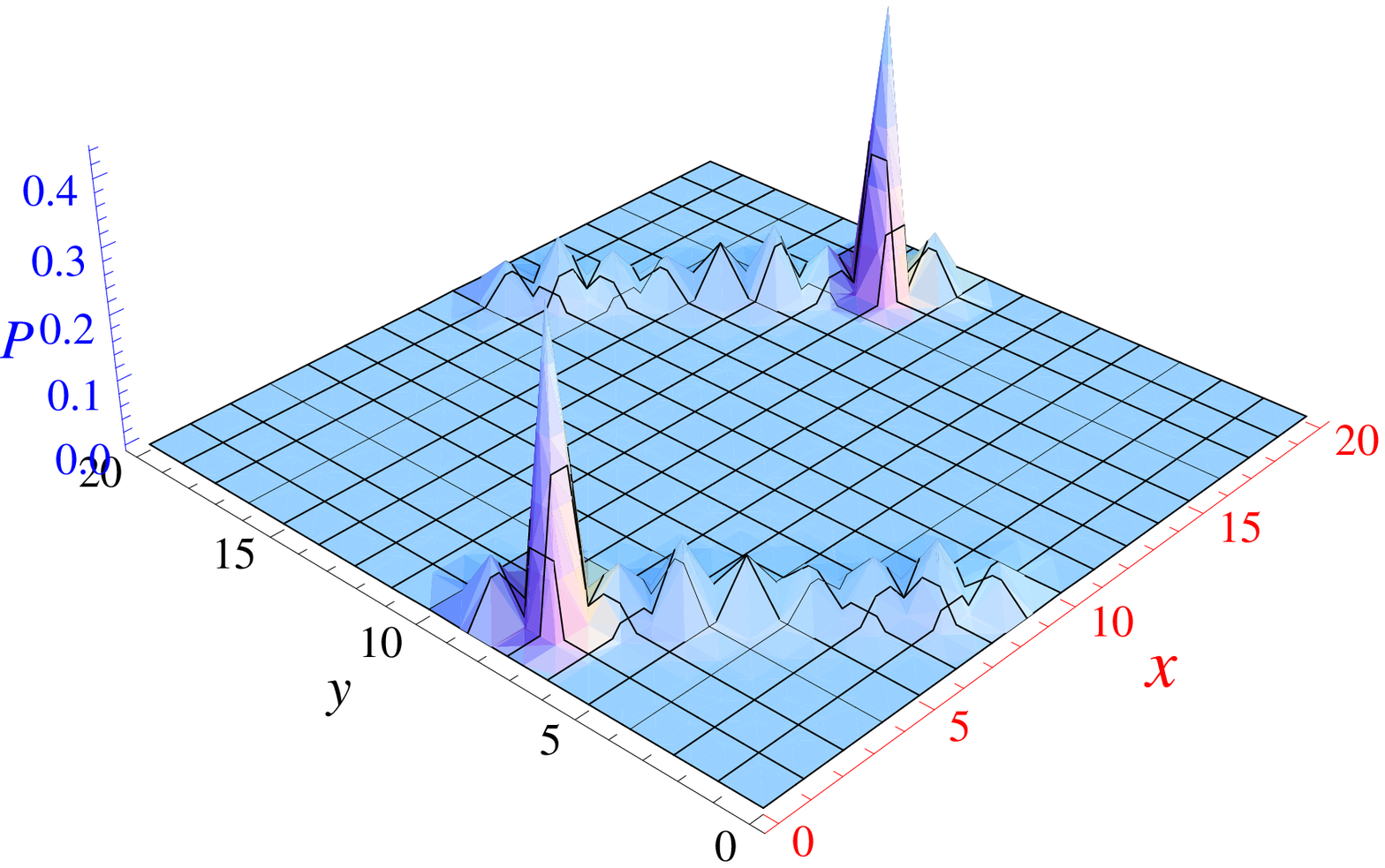}
\label{4a}}
\subfigure[]{\includegraphics[width=6.6cm]{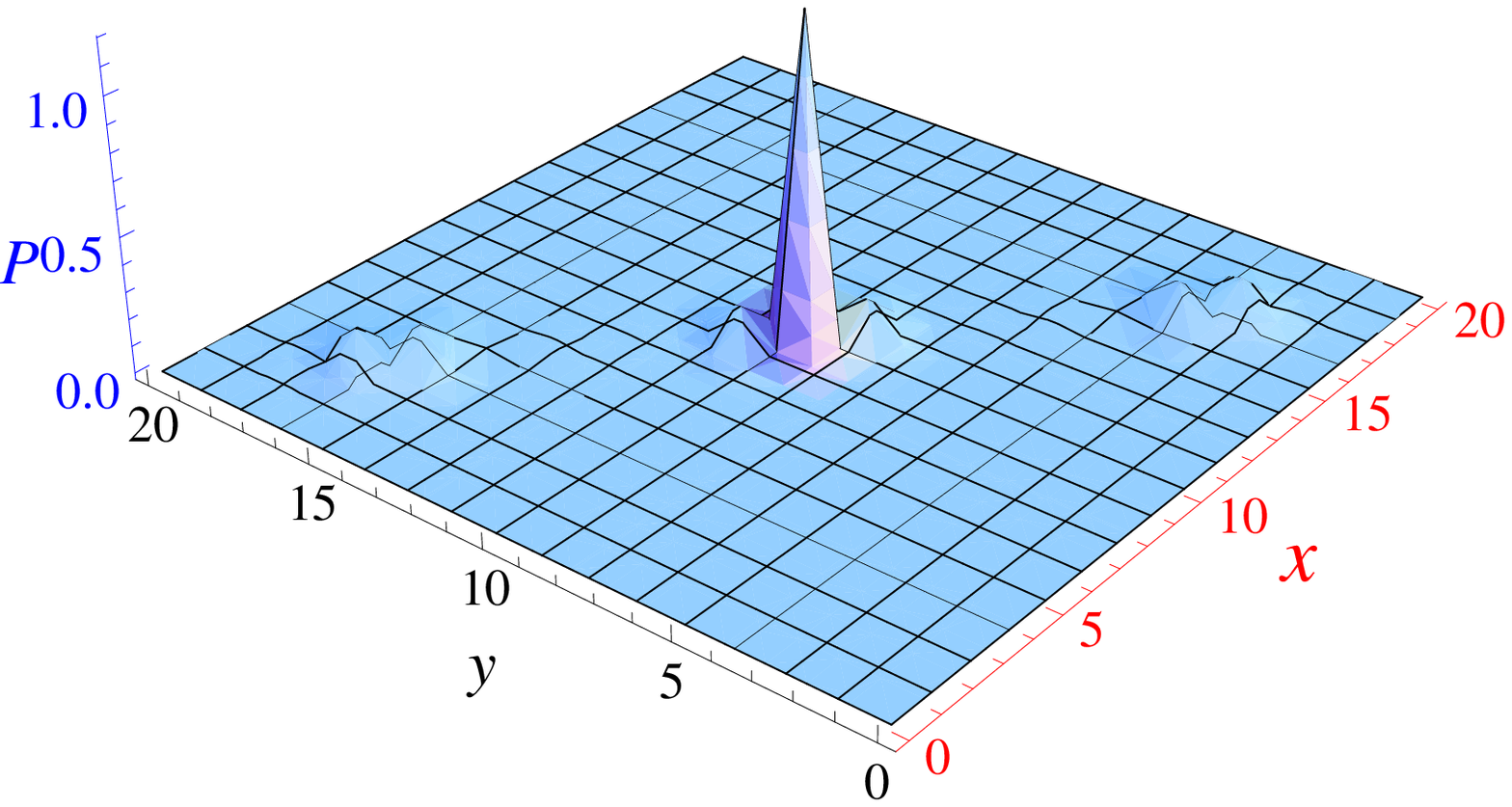}
\label{4b}}
\caption{(color online) Probability distribution of two
  distinguishable particles on a two-dimensional lattice using
  $B(\pi/4)$ as quantum coin operation. (a) Joint probability
  distribution of two particles staring at $(0, 0)$ and $(20, 20)$
  with initial states $|\downarrow \rangle$ after 10 steps. (b)
  Localization of two-particle probability distribution at the center
  of the lattice after a one time bit-flip operation on both particles
  at $t=j/2$ was introduced.}
\end{figure}
\begin{figure}[h]
\subfigure[]{\includegraphics[width=6.5cm]{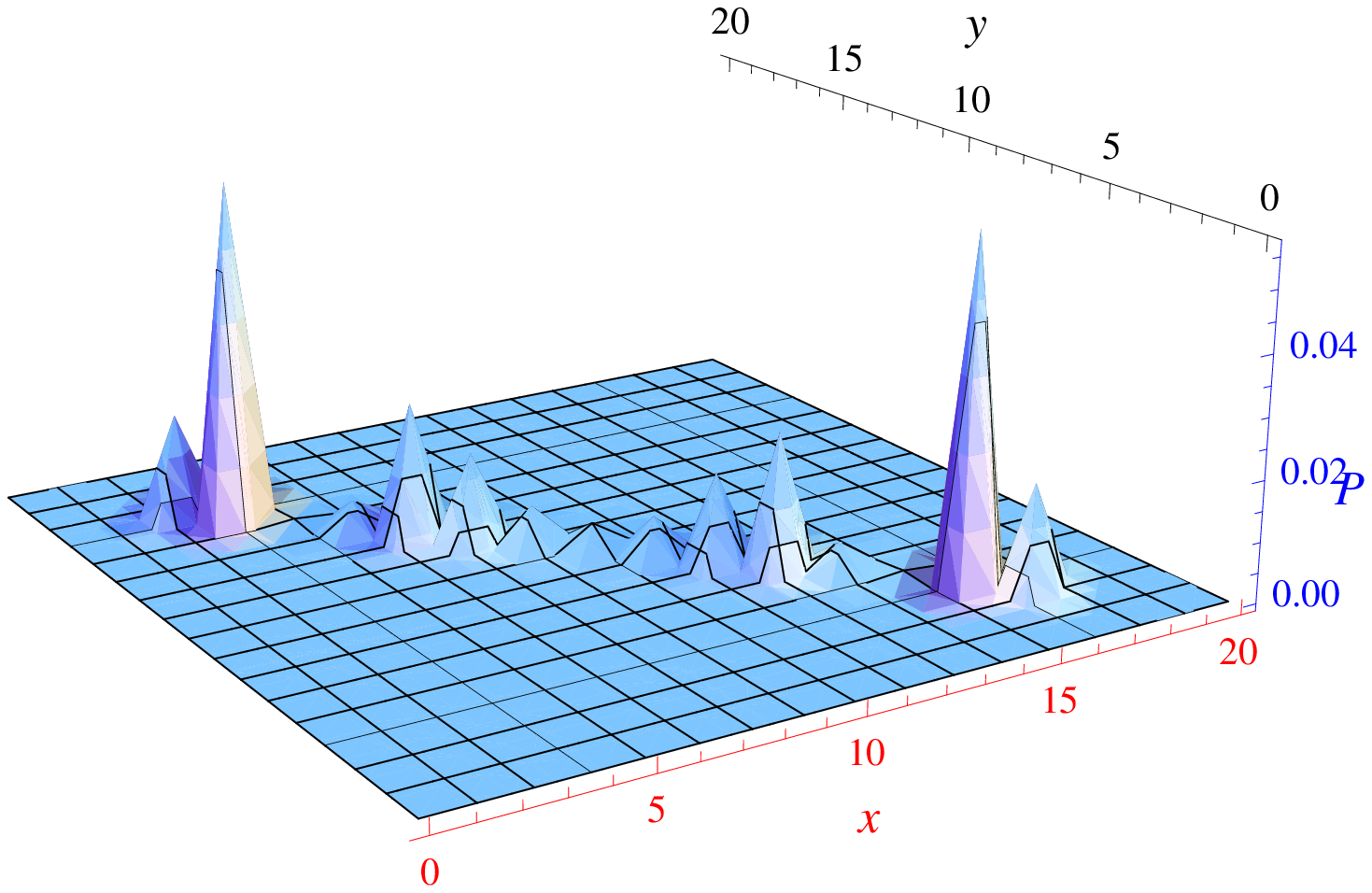}
\label{5a}}
\subfigure[]{\includegraphics[width=6.5cm]{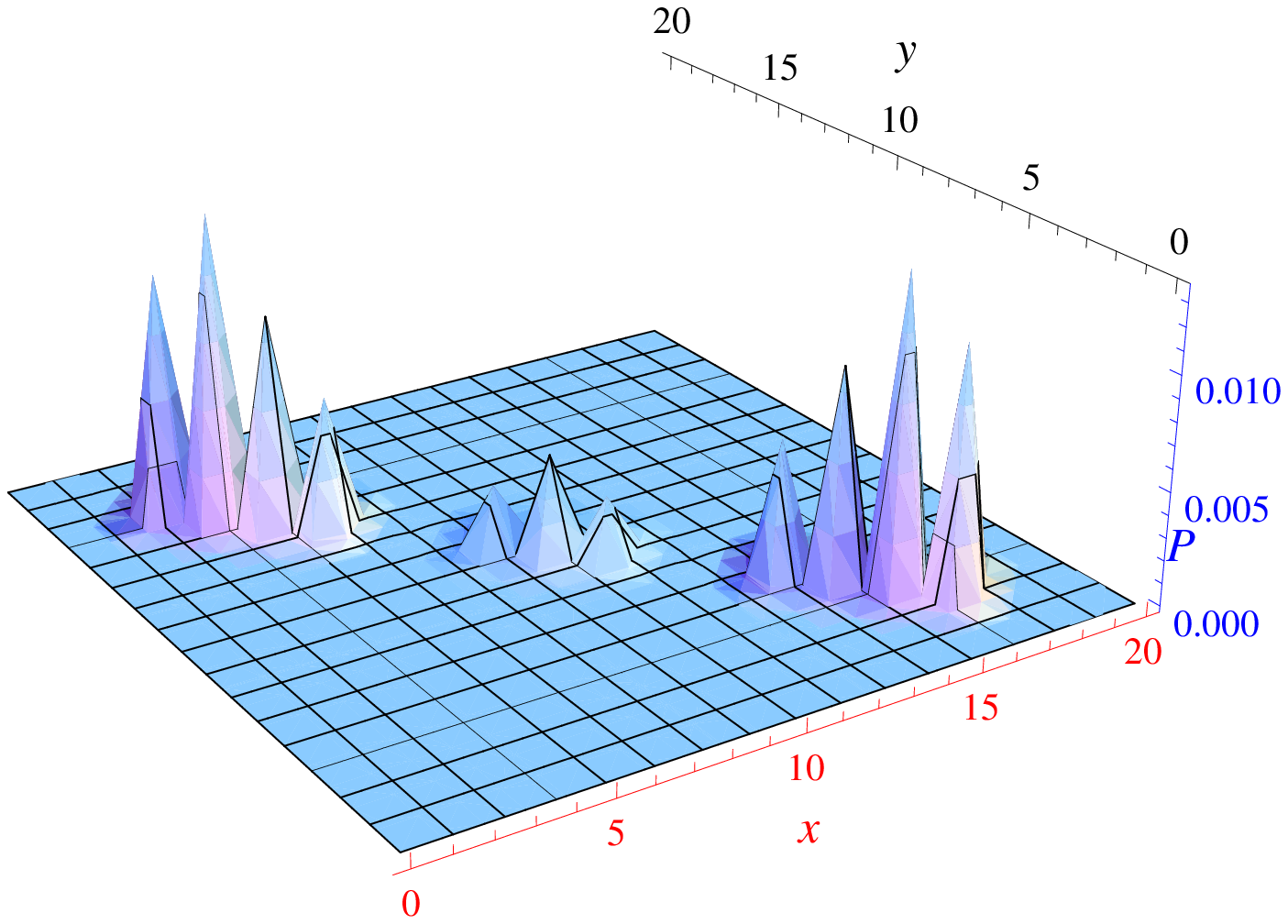}
\label{5b}}
\subfigure[]{\includegraphics[width=6.5cm]{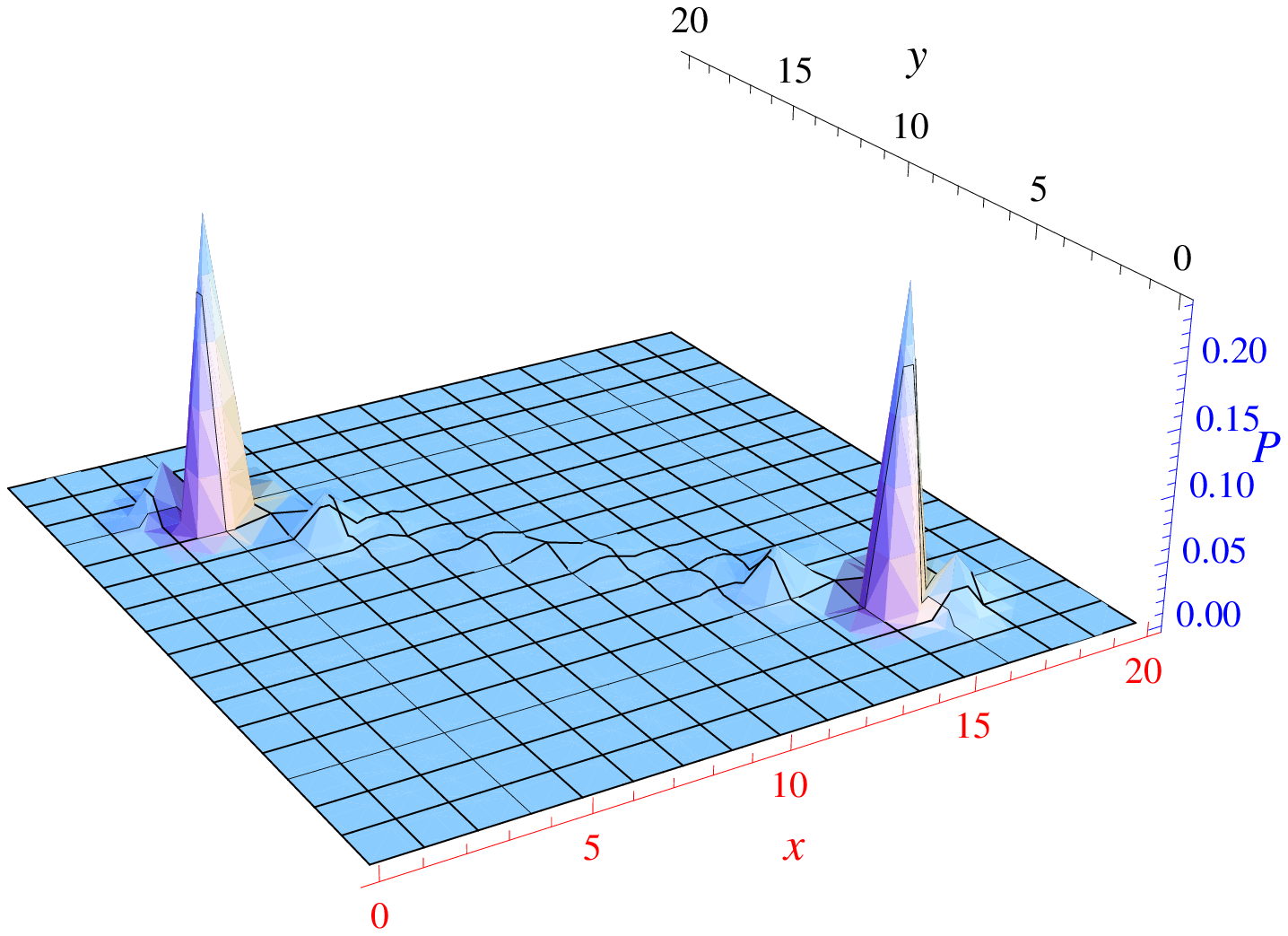}
\label{5c}}
\caption{(color online) Quantum walk on a two-dimensional lattice for
  two indistinguishable bosons initially at $(0, 0)$ and $(20, 20)$ in
  $|\downarrow\downarrow\rangle$ and interacting via $\sigma_x$ after
  20 steps using $B(\pi/4)$ as the quantum coin operation. The
  probabilities for finding the particles in the state (a)
  $|\downarrow \downarrow \rangle$ (b) $|\uparrow \uparrow \rangle$
  and (c) $|\downarrow \uparrow \rangle$ are shown at $t=20$. The
  relative height of the final distributions however can be shown to
  depend on the initial state.}
\label{intqwA}
\end{figure}
  
%===================
\section{Joint probability of two-particle quantum walk}
\label{2pqw}
%===================
Two-particle quantum walks have been studied from various
  perspectives \cite{OPS06, GFZ10, SBK11, BW11} and first experimental
  implementations have recently been reported \cite{ZKG10,
    PLM10}. Here we will discuss the probability distribution of a
  quantum walk using two distinguishable particles on a
  two-dimensional lattice and present a protocol to increase the
  meeting probability of the two particle at a particular lattice
  after a particular time. We then compare this to the quantum walk
  evolution of two indistinguishable particles which only interact at
  the end of a certain number of steps.
\par
 \begin{figure}[ht]
\includegraphics[width=6.7cm]{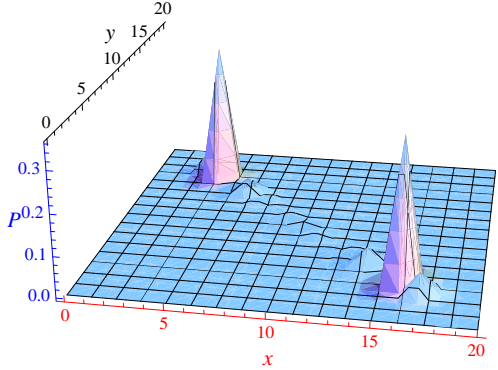}
\caption{(color online) Probability distribution of finding
  the two fermions staring at $(0, 0)$ and $(20, 20)$ with initial
  states $|\downarrow \rangle$ and interacting with $\sigma_x$ after
  20 steps of quantum walk on a two-dimensional lattice using
  $B(\pi/4)$ as quantum coin operation. The state $|\downarrow
  \uparrow \rangle$ is the only possible state for fermions.}
\label{6}
\end{figure}

To define a two-particle quantum walk we will consider a two-dimensional square lattice and label the two axis as $X$ and $Y$ such that $(x, y)$ represent a position on the lattice. We will consider two particles initially in state $|\downarrow \rangle$  at diagonally opposite points $(0, 0)$ and $(j, j)$,
\begin{equation}
|\Psi_{ins}^{2}\rangle =  \left [ | \downarrow \rangle \otimes |\psi_{0, 0}\rangle \right ] \otimes \left [ | \downarrow \rangle \otimes |\psi_{j, j}\rangle \right ],
\label{2QWstate}
\end{equation}
where $j$ is the length of the lattice, which has $j \times j$ positions. 
The shift operator for the quantum walk evolution is defined separately for both particles in such a way that they evolve towards each other,
%\begin{widetext}
\bea
S_1\equiv         \sum_{x, y}  [  |\downarrow \rangle\langle
\downarrow |\otimes|\psi_{x+1, y}\rangle\langle   \psi_{x, y}|   \nonumber \\
+  | \uparrow \rangle\langle
\uparrow|\otimes |\psi_{x, y+1}\rangle\langle \psi_{x, y}|  ] \nonumber \\
S_2\equiv         \sum_{x, y} [  |\downarrow \rangle\langle
\downarrow |\otimes|\psi_{x-1, y}\rangle\langle   \psi_{x, y}| \nonumber \\
   +  | \uparrow \rangle\langle
\uparrow|\otimes |\psi_{x, y-1}\rangle\langle \psi_{x, y}|  ].
\eea
Each step of the evolution can be implemented by $W_2(\theta) = [B(\theta) \otimes S_1]\otimes [B(\theta) \otimes S_2]$ and after $t$ steps the state is given by
\bea
\label{tpqw}
[W_2(\theta)]^t|\Psi_{ins}^{2}\rangle = \left \{  [ B(\theta) \otimes S_1]\otimes [B(\theta) \otimes S_2] \right \}^t  \nonumber \\
\times \left \{ [|\downarrow \rangle \otimes |\psi_{0,0}\rangle ] \otimes [|\downarrow \rangle \otimes |\psi_{j,j}\rangle ] \right \}.
\eea
%\end{widetext}
The two particles meet each other for the first time after $t=j$ steps and the meeting probability at each position is different for the distinguishable and indistinguishable case. 
\par
{\it Two distinguishable particles}: In this case the joint probability of the two particles at each position at the time of meeting each other is the sum of the probabilities of both individual particle.  In Fig. \ref{4a} we show this distribution for both particles after $t=j/2=10$ steps on a $20 \times 20$ lattice. The first time the two distributions overlap is at $t=20$ where they spread along the diagonal of the lattice (not shown). If, however, we introduce a one time bit-flip operation, $\sigma_x$ at t=j/2=10,
\be
[W(\theta)^{\otimes 2}]^{j/2}[\sigma_x \otimes \sigma_x][W(\theta)^{\otimes 2}]^{j/2}|\Psi_{ins}^{2}\rangle
\ee
one can see from Fig. \ref{4b} that the evolution can be reversed, which leads to localization of both the particles in the center of the lattice at time $t=j$ with a good probability. 

{\it Two indistinguishable particle}: If the two particles are indistinguishable, their probability distributions interfere when they overlap at the same position in the lattice. For bosons the allowed states at each position in the lattice are $|\downarrow
  \downarrow \rangle$, $|\uparrow \uparrow \rangle$ or $|\downarrow
  \uparrow \rangle \equiv |\downarrow \uparrow \rangle$, whereas for
  fermions these are restricted to $|\downarrow \uparrow \rangle
  \equiv |\downarrow \uparrow \rangle$. The probabilities for these
  states to be obtained at each position at time $t$ are then given
  for bosons as
\begin{widetext}
\begin{subequations}
\bea
P^{j}_{|\downarrow \downarrow \rangle}  = \frac{|\mathcal{A}^a_{j, t}|^2\cdot |\mathcal{A}^b_{j, t}|^2}{\sum_j [  |\mathcal{A}^a_{j, t}|^2 \cdot |\mathcal{A}^b_{j, t}|^2 + |\mathcal{B}^a_{j, t}|^2 \cdot |\mathcal{B}^b_{j, t}|^2 + |\mathcal{A}^a_{j, t}|^2 \cdot |\mathcal{B}^b_{j, t}|^2 + \mathcal{A}^b_{j, t}|^2 \cdot |\mathcal{B}^a_{j, t}|^2]} \\
P^{j}_{|\uparrow \uparrow \rangle}  = \frac{|\mathcal{B}^a_{j, t}|^2\cdot |\mathcal{B}^b_{j, t}|^2}{\sum_j [  |\mathcal{A}^a_{j, t}|^2 \cdot |\mathcal{A}^b_{j, t}|^2 + |\mathcal{B}^a_{j, t}|^2 \cdot |\mathcal{B}^b_{j, t}|^2 + |\mathcal{A}^a_{j, t}|^2 \cdot |\mathcal{B}^b_{j, t}|^2 + \mathcal{A}^b_{j, t}|^2 \cdot |\mathcal{B}^a_{j, t}|^2]} \\
P^{j}_{|\uparrow \downarrow \rangle}  = \frac{|\mathcal{A}^a_{j, t}|^2\cdot |\mathcal{B}^b_{j, t}|^2 + |\mathcal{A}^b_{j, t}|^2\cdot |\mathcal{B}^a_{j, t}|^2 }{\sum_j  [ |\mathcal{A}^a_{j, t}|^2 \cdot |\mathcal{A}^b_{j, t}|^2 + |\mathcal{B}^a_{j, t}|^2 \cdot |\mathcal{B}^b_{j, t}|^2 + |\mathcal{A}^a_{j, t}|^2 \cdot |\mathcal{B}^b_{j, t}|^2 + \mathcal{A}^b_{j, t}|^2 \cdot |\mathcal{B}^a_{j, t}|^2]}. 
\eea
\end{subequations}
\end{widetext}
Here $\mathcal{A}^a$ and $\mathcal{A}^b$ are the amplitudes of the
particles $a$ and $b$ to be in state $|\downarrow \rangle$, and
$\mathcal{B}^a$ and $\mathcal{B}^b$ are the amplitudes to be in the
state $|\uparrow \rangle$. We show these probabilities two particles
initially at $(0, 0)$ and $(20, 20)$ and meeting after evolving for 20
steps of walk using Eq.\,(\ref{tpqw}) in Fig.\,\ref{5a},(b) and
(c).

 If the particles are fermions the probability of finding the
two-particle in the only possible state at each positions is
\bea
P^{j}_{|\downarrow \uparrow \rangle} =\frac{ \|\mathcal{A}^a_{j,
    t}|^2\cdot |\mathcal{B}^b_{j, t}|^2 + |\mathcal{A}^b_{j,
    t}|^2\cdot |\mathcal{B}^a_{j, t}|^2}{\sum_j[\|\mathcal{A}^a_{j,
    t}|^2\cdot |\mathcal{B}^b_{j, t}|^2 + |\mathcal{A}^b_{j,
    t}|^2\cdot |\mathcal{B}^a_{j, t}|^2]},
\eea 
which is shown in Fig.\,\ref{6} for the same parameters as in the
bosonic case above. The difference to the bosonic case is clearly
visible. Using different initial states of the particle or different
coin operations during the evolution will of course alter the
probability distribution. Introducing a one time bit-flip operation half way through 
the  evolution for indistinguishable particles as we did for distinguishable particle will lead to localization of the join probability at the center (not shown).

  From this one can see that even a one time particle-particle
  interaction in an indistinguishable many-particle quantum walk can
  result in different probability distributions which might be useful
  for applications in quantum information and other fundamental
  quantum mechanical experiments.  With the possibility of increasing
  the number of steps, the number of particles and the number of time
  the particle-particle interaction is introduced, the evolution gets
  even more interesting and complicated, but becomes computationally
  difficult. Recently,for the case of two atoms in an optical lattice
  performing a quantum walk with interactions via cold collisions the
  appearance of a bound state has been predicted \cite{ AAM11}, which
  gives scope for further exploration of the dynamics using our
  approach for many-particle system by introducing interactions at
  regular intervals. 

%===================
\section{Conclusion}
\label{conc}
%===================
 We have presented a number examples of quantum walk dynamics of
  many-particle system with different initial states of the
  particles. Though the distinguishable many-particle quantum walk
  dynamics does not involve many-particle interference during the
  evolution we have shown that it can be effectively used to
  separate the eigenstates of the particles position space and group
  them together. We have also presented an example of two-particle quantum walk
  dynamics with defined interaction that can lead to localization of two distinguishable particles at the
  center if they start their walk from opposite ends of the lattice. Extending this scheme to 
  indistinguishable boson and fermion pairs results in the different probabilities for finding the two
  particles in the allowed combination of states. Recent experimental
  developments in implementing quantum walks and using quantum walk
  models to simulate and understand some of the dynamics process in
  nature suggests that collective dynamics of many-particle system
  will very useful for further studies.

%===================


\begin{thebibliography}{99}
%===================

\bibitem{Ria58}  G. V. Riazanov, Sov. Phys. JETP {\bf 6} 1107 (1958).

\bibitem{FH65} R. P. Feynman and A.R. Hibbs, {\it Quantum Mechanics and Path Integrals} (McGraw-Hill, New York, 1965).

\bibitem {ADZ93} Y. Aharonov, L. Davidovich and N. Zagury, Phys. Rev. A {\bf 48}, 1687, (1993).

\bibitem{DM96} D.  A. Meyer,  J. Stat. Phys. {\bf 85},  551 (1996).

\bibitem{FG98} E. Farhi and S. Gutmann, Phys.Rev. A {\bf 58}, 915 (1998).
\bibitem{ABN01} A. Ambainis, E. Bach, A. Nayak, A. Vishwanath and J. Watrous, {\it Proceeding of the 33rd ACM Symposium on Theory of Computing} (ACM Press, New York, 2001), p.60.

\bibitem{NV01} A. Nayak and A. Vishwanath, DIMACS Technical Report, No. 2000-43  (2001) ; arXiv:quant-ph/0010117.

\bibitem{Amb03} A. Ambainis, Int. Journal of Quantum Information, {\bf 1}, No.4, 507-518 (2003).

\bibitem{CCD03} A. M. Childs, R. Cleve, E. Deotto, E. Farhi, S. Gutmann and D. A. Spielman, in {\it Proceedings of the 35th ACM Symposium on Theory of Computing} (ACM Press, New York, 2003), p.59.

\bibitem{SKB03} N. Shenvi, J. Kempe and K. Birgitta Whaley, Phys. Rev. A {\bf 67}, 052307, (2003).

\bibitem{AKR05} A. Ambainis, J. Kempe, and A. Rivosh, {\it Proceedings of ACM-SIAM
Symp. on Discrete Algorithms (SODA)},  (AMC Press, New York, 2005), pp.1099-1108.

\bibitem{CL08} C. M. Chandrashekar and R. Laflamme,  Phys. Rev. A  {\bf 78},  022314 (2008).

\bibitem{OKA05} T. Oka, N. Konno, R. Arita, and H. Aoki, Phys. Rev. Lett. {\bf 94}, 100602 (2005). 

\bibitem{ECR07} G. S. Engel {\it et. al.}, Nature {\bf 446}, 782-786 (2007).  

\bibitem{MRL08} M. Mohseni, P. Rebentrost, S. Lloyd, A. Aspuru-Guzik, J. Chem. Phys. {\bf 129}, 174106 (2008).

\bibitem{CGB10} C. M. Chandrashekar, Sandeep K Goyal, and Subhashish Banerjee, arXiv:1005.3785 (2010).

\bibitem{KRB10} T. Kitagawa, M. S. Rudner, E. Berg, and E. Demler, Phys. Rev. A 82, 033429 (2010).


\bibitem{DLX03} J. Du, H. Li, X. Xu, M. Shi, J. Wu, X. Zhou, and R. Han, Phys. Rev. A {\bf 67}, 042316 (2003)

\bibitem{RLB05} C. A.  Ryan, M.  Laforest, J. C. Boileau, and R. Laflamme, Phys. Rev. A
{\bf 72}, 062317 (2005).

\bibitem{PLP08} H. B. Perets, Y. Lahini, F. Pozzi, M. Sorel, R. Morandotti, and Y. Silberberg,  Phys. Rev. Lett. {\bf 100}, 170506 (2008).

\bibitem{SMS09} H. Schmitz, R. Matjeschk, Ch. Schneider, J. Glueckert, M. Enderlein, T. Huber, and T. Schaetz, Phys. Rev. Lett. {\bf 103}, 090504 (2009).

\bibitem{ZKG10} F. Zahringer, G. Kirchmair, R. Gerritsma, E. Solano, R. Blatt, and C. F. Roos, Phys. Rev. Lett. {\bf 104}, 100503 (2010). 

\bibitem{KFC09} K. Karski, L. Foster, J.-M. Choi, A. Steffen, W. Alt, D. Meschede, and A. Widera, Science  {\bf 325}, 174 (2009).

\bibitem{SCP10} A. Schreiber, K. N. Cassemiro, V. Potocek, A. Gabris, P. Mosley, E. Andersson, I. Jex, and Ch. Silberhorn, Phys. Rev. Lett., {\bf 104}, 05502 (2010).

\bibitem{BFL10}  M. A. Broome, A. Fedrizzi, B. P. Lanyon, I. Kassal, A. Aspuru-Guzik, and A. G. White. Phys. Rev. Lett. {\bf 104}, 153602 (2010).

\bibitem{PLM10} Alberto Peruzzo, Mirko Lobino, Jonathan C. F. Matthews, Nobuyuki Matsuda, Alberto Politi, Konstantinos Poulios, Xiao-Qi Zhou, Yoav Lahini, Nur Ismail, Kerstin Wörhoff4, Yaron Bromberg, Yaron Silberberg, Mark G. Thompson and Jeremy L. OBrien, Science, {\bf 329},  1500-1503 (2010). 

\bibitem{OBB11} J. O. Owens, M. A. Broome, D. N. Biggerstaff, M. E. Goggin, A. Fedrizzi, T. Linjordet, M. Ams, G. D. Marshall, J. Twamley, M. J. Withford and A. G. White, New J. Phys. {\bf 13}, 075003 (2011).

\bibitem{Kon02} N. Konno, Quantum Information Processing, {\bf 1}, Issue 5, pp.345-354 (2002).

\bibitem{CSL08} C. M. Chandrashekar, R. Srikanth, and R. Laflamme,  Phys. Rev. A {\bf 77}, 032326 (2008).

\bibitem{MTM11} Klaus Mayer, Malte C. Tichy, Florian Mintert, Thomas Konrad, and Andreas Buchleitner, Phys. Rev. A {\bf 83}, 062307 (2011).

\bibitem{RSS11} Peter. P. Rohde, Andreas Schreiber, Martin Stefanak, Igor Jex and Christine Silberhorn, New J. Phys. {\bf 13}, 013001 (2011).

\bibitem{GC10} Sandeep K Goyal and C. M. Chandrashekar, J. Phys. A: Math. Theor. {\bf 43}, 235303 (2010).

\bibitem{MGW03} O. Mandel, M. Greiner, A. Widera, T. Rom,  T.W. H\"ansch, and I. Bloch, Phys. Rev. Lett. {\bf 91}, 010407 (2003).

\bibitem{DDL03} L.-M. Duan, E. Demler, M. D. Lukin, Phys. Rev. Lett. {\bf 91}, 090402 (2003).

\bibitem{Jak04} D. Jaksch, Contemporary Physics, {\bf 45} No. 5, 367-381 (2004).

\bibitem{SKJ06} M. Stefanak, T. Kiss, I. Jex, and B. Mohring, J. Phys. A : Math. Gen. {\bf 39} 14965-14983 (2006).

\bibitem{KRS03} P. L. Knight, E. Roldan, and J. E. Sipe. Quantum walk on the line as an
interference phenomenon. Phys. Rev. A, {\bf 68} 020301(R) (2003).

\bibitem{OPS06} Y. Omar, N. Paunkovic, L. Sheridan, and S. Bose, Phys. Rev. A {\bf 74} 042304 (2006).

\bibitem{GFZ10} John King Gamble, Mark Friesen, Dong Zhou, Robert Joynt, and S. N. Coppersmith, Phys. Rev. A {\bf 81} 052313 (2010).

\bibitem{SBK11} M. Stefanak , S. M.  Barnett, B. Kollar, T. Kiss and I. Jex, New Journal of Physics {\bf 13} 033029  (2011).

\bibitem{BW11} Scott D. Berry and Jingbo B. Wang, Phys. Rev. A {\bf 83} 042317 (2011).


\bibitem{Rom10} Alejandro Romanelli, Phys. Rev. A, {\bf 81} 062349 (2011).

\bibitem{AAM11} Andre Ahlbrecht, Andrea Alberti, Dieter Meschede, Volkher B. Scholz, Albert H. Werner, and Reinhard F. Werner, arXiv:1105.1051v1 (2011).



                                  
\end{thebibliography}
\end{document}